\documentclass[runningheads]{llncs}
\usepackage[T1]{fontenc}
\usepackage{graphicx}
\usepackage{array}
\usepackage{booktabs}

\usepackage{tikz,xcolor,hyperref,xurl}
\definecolor{lime}{HTML}{A6CE39}
\DeclareRobustCommand{\orcidicon}{%
	\begin{tikzpicture}
	\draw[lime, fill=lime] (0,0) 
	circle [radius=0.16] 
	node[white] {{\fontfamily{qag}\selectfont \tiny ID}};
	\draw[white, fill=white] (-0.0625,0.095) 
	circle [radius=0.007];
	\end{tikzpicture}
	\hspace{-2mm}
}
\foreach \x in {A, ..., Z}{%
	\expandafter\xdef\csname orcid\x\endcsname{\noexpand\href{https://orcid.org/\csname orcidauthor\x\endcsname}{\noexpand\orcidicon}}
}

\begin{document}
\title{User-Reported Misinformation Exposure Across Social Media Platforms}
\titlerunning{Reported Misinformation Exposure Across Platforms}
%
\author{Catherine King\inst{1}\orcidA \and
Lynnette Hui Xian Ng\inst{1}\orcidB \and
Kathleen M. Carley\inst{1}\orcidC}
\authorrunning{King et al.}
%
\institute{Carnegie Mellon University, Pittsburgh, PA 15213, USA\\
\email{cking2@cs.cmu.edu, lynnetteng@cmu.edu, kathleen.carley@cs.cmu.edu}}

%
\maketitle              
\begin{abstract}
In this study, we surveyed users for their perception of misinformation exposure across social media platforms. Such perceived exposure is important because individuals' beliefs about how often they encounter false information can shape their trust in institutions, platforms, and even their friends. In a survey of 1,010 United States residents, we found that perceived exposure to misinformation varies substantially across platforms and is only moderately correlated with the frequency of platform use. A much larger percentage of participants also reported being exposed to misinformation from the public feed than from known contacts. Based on these results, we propose governance strategies across three categories of platform types: discovery, interpersonal, and discourse. This work offers insight into users' perceptions of social media misinformation and a corresponding research agenda for governance.

\end{abstract}
\section{Introduction}
Social media has become one of the primary channels for discovering news \cite{oeldorf2022unavoidable}, and at the same time, it is one of the main vectors of misinformation, especially during periods of social uncertainty, like geopolitical crises \cite{madrid2024geopolitics}. Much of the existing literature focuses on observable diffusion dynamics and engagement with misinformation, rather than users' subjective perceptions of misinformation exposure itself and how this varies across different social media environments \cite{ng2024exploring}. Perceived exposure is important because individuals' beliefs about how often they encounter false information can shape their trust in institutions, platforms, and even their friends \cite{southwell2018misinformation}.

Misinformation exposure varies across platforms \cite{moslehDivergentPatternsEngagement2024}. Different platform features and recommendation mechanisms shape how users encounter information. Platforms centered around public broadcasting, like X, may facilitate rapid amplification of misleading narratives, while entertainment-oriented platforms like TikTok expose users through personalized recommendations, and professional platforms like LinkedIn have less misinformation \cite{moslehDivergentPatternsEngagement2024}. As platforms navigate the web of information governance and moderation, it is important to understand how users perceive their exposure to misinformation across platforms. Knowing how often users encounter misinformation and on which platforms, from a user perspective, is important for designing platforms that help users better assess the validity of online information. 
To understand the perceptions of misinformation exposure across social media platforms, we surveyed 1,010 US residents about their experiences with misinformation. We investigate the following:

\textbf{RQ1}. Which social media platforms do people report being exposed to or posting misinformation on? 

\textbf{RQ2}. Is platform usage related to misinformation exposure?

\section{Related Work}
Research on misinformation has extensively examined how misleading information spreads through social networks, causing paranoid thinking as in the Kennedy assassination \cite{gagne2022thinking} or the rejection of critical vaccines \cite{ng2024exploring}. Misinformation spreads rapidly across social media ecosystems due to network amplification, emotional engagement, and platform algorithms \cite{aimeur2023fake,ng2024exploring}. Prior studies show that engagement with low-quality or partisan news varies across platforms, with some disproportionately amplifying misleading or politically extreme content \cite{moslehDivergentPatternsEngagement2024}. Additionally, platform moderation and governance shape the visibility and perception of misinformation \cite{schaffnerCommunityGuidelinesMake2024}. Platforms have employed different approaches to moderation, ranging from community notes on X, community moderation on Reddit, and platform-wide removal of harmful accounts on Facebook \cite{kumar2023governance}.

While misinformation exposure is less studied, a growing body of work suggests that it differs across platforms due to affordances like how connections are formed (profile-based versus content-based) and message types (personalized for users versus public broadcasts). These structural differences create distinct pathways through which information spreads and is received on platforms \cite{ng2021does}. Further, users engage with different platforms to fulfill diverse sets of information and social needs \cite{voorveldEngagementSocialMedia2018,whiting2013people,zhuSocialMediaHuman2015}, suggesting that users may interpret and respond differently to information depending on the platform context \cite{moslehDivergentPatternsEngagement2024}. 

\section{Survey Methodology}
To understand people's perceptions of misinformation on social media, we surveyed 1,010 participants between July and August 2024 through Qualtrics. These participants were screened for the following criteria: use of social media at least once a week, US residency, and age (18 or older). We employed several methods to recruit relevant participants and maintain high data quality \cite{king2025scirep,king2026jots}. The survey was approved by Carnegie Mellon University's IRB under the number “STUDY2022\_00000143”. Informed consent was obtained from the participants, who were paid the equivalent of \$10/hour. In the survey, the participants were asked the following questions related to their perceived misinformation exposure. We note that there were additional survey questions beyond these research questions that were analyzed in other work \cite{king2025scirep,king2026jots} and therefore not listed\footnote{See \url{https://doi.org/10.1184/R1/27264813} for a complete copy of the survey.}.

\begin{itemize}
    \item \textbf{Platform usage}: Participants were asked how often they visited the top 11 most popular social media platforms \cite{auxierSocialMediaUse2021}, or a platform not listed \textit{\{Multiple times a day, Daily, Weekly, Monthly, Less often than monthly, Never\}}. 
    \item \textbf{Exposure to misinformation}: Participants were asked whether their social media contacts had ever posted something they believed to be misinformation, on which platforms they had been exposed to it, and whether it came from known or unknown contacts. 
    \item \textbf{Posting misinformation}: Participants were asked if they had ever intentionally or unintentionally posted misinformation, and on which platforms.
    \item \textbf{Demographic Question}: Participants answered demographic questions including age, gender, education, income, and political ideology. 
\end{itemize}

\section{Results and Analysis}
\begin{figure}[htp]
\centering
\includegraphics[width=.9\textwidth]{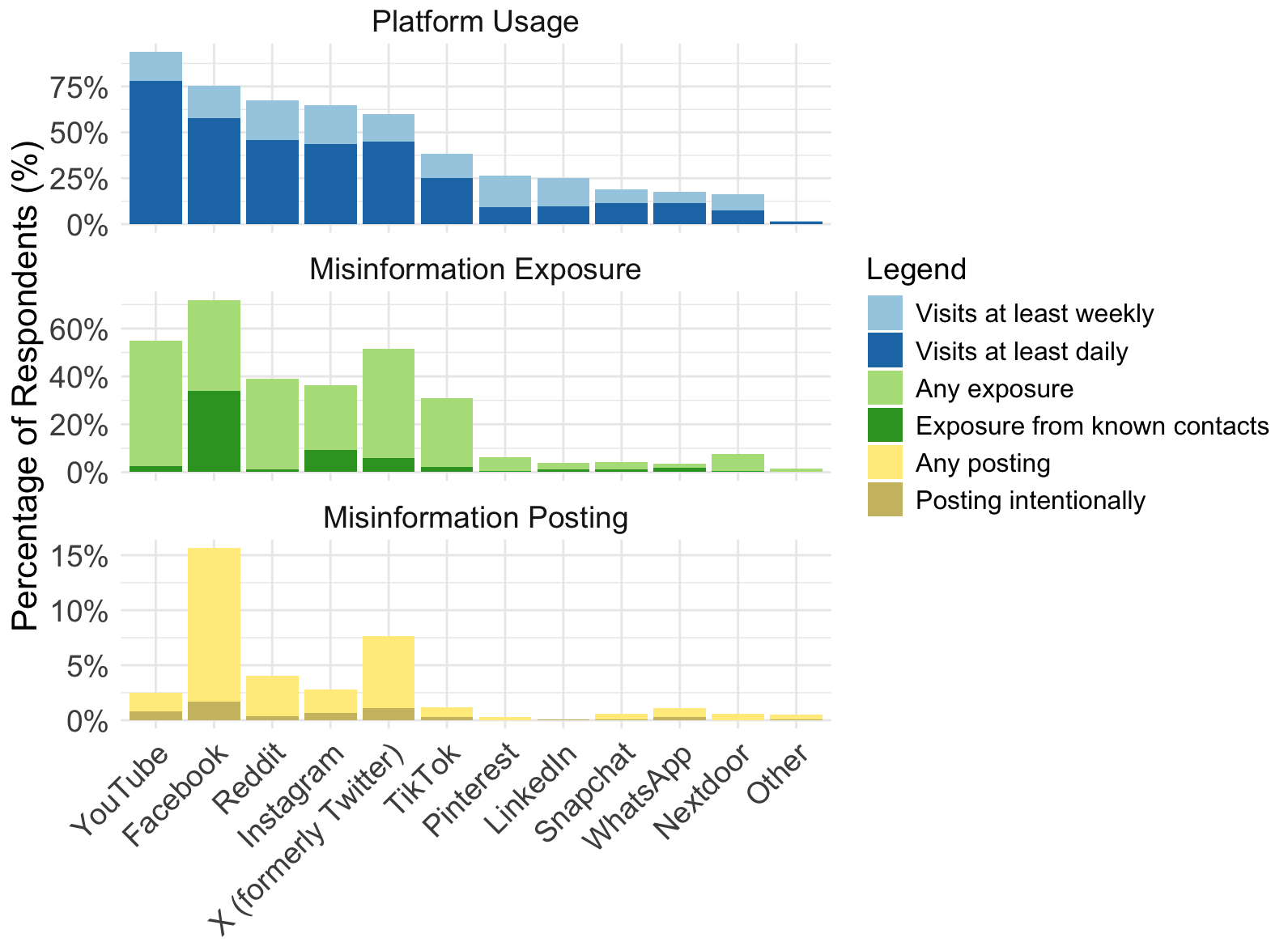}
\caption{The percentage of participants who reported encountering or posting misinformation on a platform and their reported platform usage.} \label{fig:updated}
\end{figure}

\autoref{fig:updated} reports the most visited platforms and participants' reported exposure to and posting of misinformation. The most visited platforms were YouTube, Facebook, Reddit, X, and Instagram, with over half of participants visiting them weekly. However, usage did not fully align with reported misinformation exposure. Facebook had the most reported exposure, followed by YouTube, X, Reddit, Instagram, and TikTok. 

The platforms also differed substantially on the source of the misinformation, with more participants encountering it on their public feeds than from known contacts. Among those who saw misinformation on each platform, WhatsApp (50\%), Facebook (47.4\%), LinkedIn (30\%), Snapchat (27.9\%), and Instagram (25.1\%) had the highest percentages of respondents who reported seeing misinformation from a known contact. On all other platforms, at least 85\% of participants who encountered misinformation saw it only from unknown contacts.

\autoref{fig:updated} further shows the distribution of participants who believed they had posted misinformation. About 25\% believed they had done so unintentionally, compared to around 3\% who admitted to doing so intentionally. Most of these cases occurred on Facebook, followed by X. This suggests that participants generally do not deliberately spread false information.
Interestingly, YouTube does not appear in the top three. This mirrors the pattern of self-reported misinformation exposure, suggesting that platforms associated with greater misinformation visibility may also facilitate greater unintentional sharing. 

These findings suggest that perceived exposure is not simply a function of platform use. YouTube was the most used platform, but Facebook was where users reported the most exposure. This may reflect platform design differences and context-specific user motivations \cite{pelletier2020one,voorveldEngagementSocialMedia2018}. Facebook and X are centered around opinion-based content and public discussion, which may increase exposure to contested narratives, while more entertainment-focused platforms like YouTube and TikTok show lower exposure, and false information may be perceived differently, e.g. as part of the entertainment value. The high reported exposure via WhatsApp and Snapchat highlights the role closed platforms play in shaping these perceptions \cite{butler2023misinformation}.
Table \ref{tab:correlation_matrix} shows a moderate, positive correlation between weekly platform usage and self-reported misinformation exposure, but only weak positive correlations between weekly usage and self-reported posting, as well as between exposure and posting. Overall, these patterns suggest that platform architecture and interaction mechanisms, not just usage frequency alone, shape how users perceive their exposure to and propagation of misinformation. 

\begin{table}[h]
\centering
\footnotesize
\caption{Correlation matrix between weekly usage, reported misinformation exposure, and self-reported misinformation posting. * indicates significance at $p<0.01$.}
\label{tab:correlation_matrix}
\small
\begin{tabular}{lccc}
\toprule
 & \textbf{Usage} & \textbf{Exposure} & \textbf{Posting} \\
\midrule
\textbf{Usage}      & -     &  & \\
\textbf{Exposure}  & 0.563* & -     &  \\
\textbf{Posting}   & 0.183* & 0.234* & -     \\
\bottomrule
\end{tabular}
\end{table}

Next, we fit three mixed-effects logistic regression models (\texttt{lme4} package in R) with random intercepts for users to assess whether the frequency of platform use (never to multiple times a day, coded 0-5) predicts self-reported misinformation exposure (Model 1) and posting (Model 2). Model 3 uses both frequency and exposure to predict posting. Table \ref{tab:combined_models} presents the model results as odds ratios (ORs). For each outcome, the demographic covariates (age, gender, income, education, political ideology) were retained only if they improved the model fit. 

\begin{table}[htbp]
\centering
\footnotesize
\renewcommand{\arraystretch}{0.85}
\caption{Fixed effects (Models 2--3) and derived platform-specific frequency slopes (Model 1) from mixed-effects logistic regression models predicting misinformation exposure, posting, and the exposure--posting relationship. N = 1,010.}
\label{tab:combined_models}
\begin{tabular}{l>{\centering\arraybackslash}p{2.9cm}>{\centering\arraybackslash}p{3.15cm}>{\centering\arraybackslash}p{3.15cm}}
\toprule
\textbf{Predictor} & \shortstack{\textbf{Model 1:}\\\textbf{Exposure}\\\textbf{OR [95\% CI]}} & \shortstack{\textbf{Model 2:}\\\textbf{Posting}\\\textbf{OR [95\% CI]}} & \shortstack{\textbf{Model 3:}\\\textbf{Exp.$\rightarrow$Post.}\\\textbf{OR [95\% CI]}} \\
\midrule
Frequency of use & 2.41 [2.00, 2.91]$^{***}$ & 2.55 [2.22, 2.93]$^{***}$ & 2.17 [1.86, 2.54]$^{***}$ \\
Exposure & --- & --- & 5.63 [3.48, 9.10]$^{***}$ \\
\addlinespace
\multicolumn{4}{l}{\textit{Gender (ref.\ Female)}} \\
\quad Male       & 1.43 [1.10, 1.87]$^{**}$   & --- & --- \\
\quad Other      & 1.51 [0.64, 3.55]     & --- & --- \\
Political ideology & 0.89 [0.79, 0.99]$^{*}$ & --- & --- \\
\addlinespace
\multicolumn{4}{l}{\textit{Platform effect (ref.\ YouTube)}\textsuperscript{a}} \\
\quad Facebook   & 3.75 [3.27, 4.31]$^{***}$ & 26.12 [14.74, 46.26]$^{***}$ & 17.63 [10.01, 31.02]$^{***}$ \\
\quad Instagram  & 3.18 [2.75, 3.68]$^{*}$ & 2.70 [1.35, 5.41]$^{**}$     & 2.26 [1.11, 4.58]$^{*}$      \\
\quad LinkedIn   & 3.02 [2.23, 4.09] & 0.31 [0.04, 2.65]       & 0.37 [0.03, 4.68]       \\
\quad Nextdoor   & 4.40 [3.48, 5.55]$^{***}$ & 1.98 [0.60, 6.59]       & 2.46 [0.73, 8.24]       \\
\quad Other      & ---\textsuperscript{b} & 7.78 [2.16, 28.02]$^{**}$   & 7.24 [1.86, 28.12]$^{**}$    \\
\quad Pinterest  & 3.49 [2.68, 4.56]$^{*}$ & 0.67 [0.15, 3.01]       & 0.97 [0.22, 4.26]       \\
\quad Reddit     & 3.64 [3.05, 4.33]$^{**}$ & 3.85 [2.00, 7.42]$^{***}$    & 3.77 [1.98, 7.19]$^{***}$    \\
\quad Snapchat   & 2.76 [2.19, 3.50] & 1.41 [0.44, 4.57]       & 2.12 [0.66, 6.76]       \\
\quad TikTok     & 3.67 [3.22, 4.19]$^{***}$ & 1.73 [0.72, 4.20]       & 1.22 [0.50, 2.98]       \\
\quad Twitter/X  & 3.53 [3.11, 4.00]$^{***}$ & 11.17 [6.10, 20.45]$^{***}$  & 8.36 [4.59, 15.24]$^{***}$   \\
\quad WhatsApp   & 2.78 [2.21, 3.50] & 2.45 [0.94, 6.38]       & 3.65 [1.35, 9.85]$^{*}$      \\
\midrule
\textit{N} (observations) & 11{,}110 & 12{,}120 & 12{,}120 \\
R\textsuperscript{2} Marg. / Cond. & 0.613 / 0.801 & 0.403 / 0.747 & 0.400 / 0.773 \\
RMSE & 0.26 & 0.13 & 0.12 \\
\bottomrule
\end{tabular}
\raggedright
\footnotesize
\textit{Note.} Reference categories: YouTube (platform), Female (gender), Ideology (very liberal = 0 to very conservative = 4). $^{***}p<.001$, $^{**}p<.01$, $^{*}p<.05$. \\
\textsuperscript{a} Model 1 platform estimates reflect the combined effect of frequency of use within each platform (main platform effect + frequency $\times$ platform interaction), averaged over gender, derived via the \texttt{emmeans} R package; significance stars indicate whether each platform's slope differs from YouTube's. The platform estimates for Models 2 and 3 are raw fixed-effect coefficients (no interaction term in these models).\\
\textsuperscript{b} ``Other'' platform excluded from Model 1 due to complete separation (exposure perfectly predicted by frequency within this category).\\

\end{table}

Across all three models, the frequency of platform usage was a strong, independent predictor of both self-reported misinformation exposure and posting, with similar effect sizes across outcomes (ORs 2.17–2.55). Although gender and political ideology were associated with self-reported exposure, no demographic variables were associated with self-reported posting, whether modeled by platform use frequency (Model 2) or prior exposure (Model 3). Reported posting varied substantially by platform, with much higher odds on Facebook, Twitter, Reddit, and Instagram (Models 2 and 3). Exposure was one of the strongest predictors of posting (OR = 5.63), indicating that once someone encounters misinformation, the frequency of use and the platforms used best explain whether they share it, rather than their demographic background. These findings suggest a two-stage process: demographics influence who claims misinformation exposure, while engagement with the platforms determines who posts it afterward.

\section{Implications for Platform Governance}
\begin{figure}[htp]
\centering
\includegraphics[width=\textwidth]{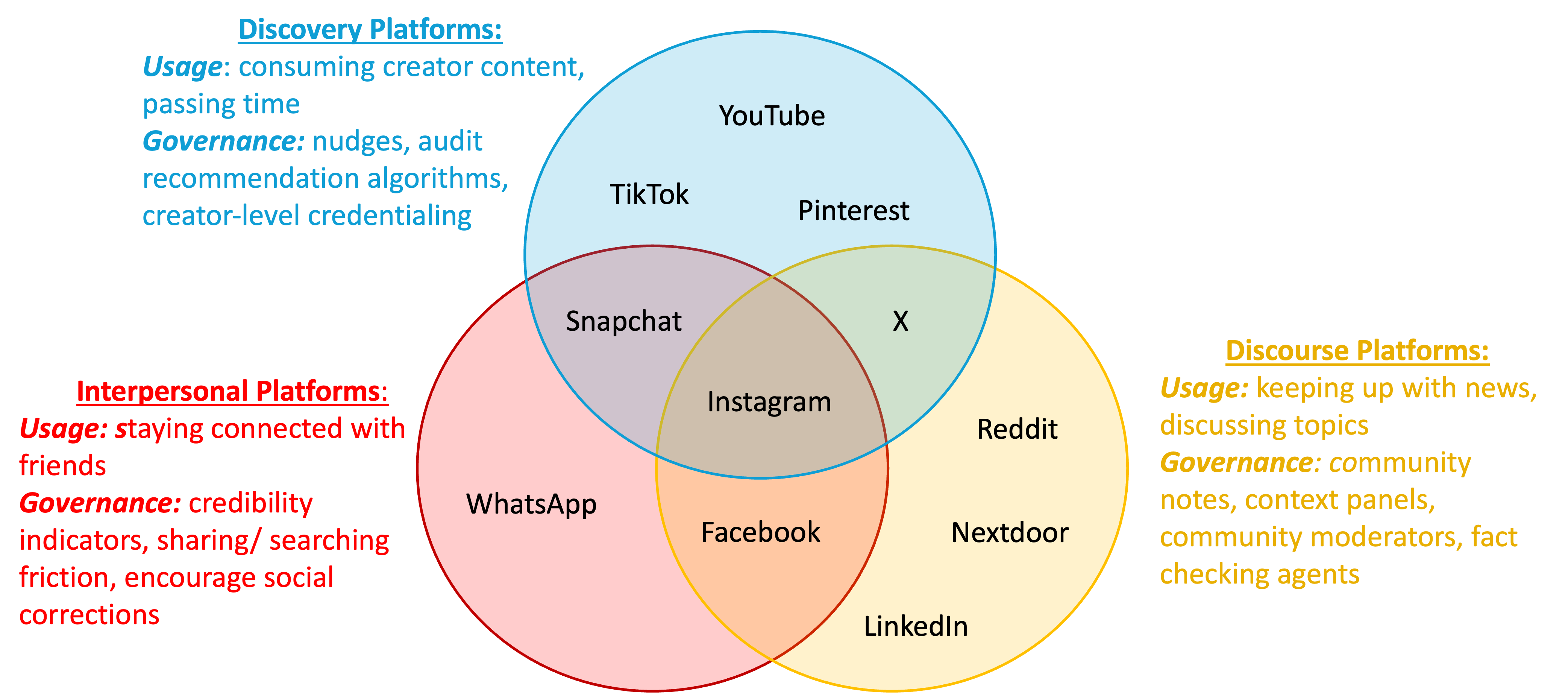}
\caption{Social media categories, primary uses, and proposed governance strategies.} \label{fig:diagram}
\end{figure}

Together, these results show that self-reported exposure to and posting of misinformation are shaped by platform-specific factors, not just demographics. This finding is consistent with previous research showing users have different expectations and reasons for using different platforms \cite{alhabash2024so,pelletier2020one,voorveldEngagementSocialMedia2018,whiting2013people}. Governance frameworks should therefore be tailored to each platform and continuously updated as its use evolves, rather than applying a single approach. Combining aspects of platform structure with a uses-and-gratifications perspective on why people consume social media \cite{whiting2013people,zhuSocialMediaHuman2015}, we propose three main categories of social media platforms, presented in \autoref{fig:diagram}: \textbf{interpersonal}, \textbf{discourse}, and \textbf{discovery}. 

\subsection{Interpersonal Platforms}
Interpersonal platforms, like WhatsApp \cite{rossini2021dsyfunctional}, Snapchat \cite{voorveldEngagementSocialMedia2018}, Facebook \cite{rossini2021dsyfunctional}, and Instagram \cite{lu2022insta}, are built around interactions with known contacts and have been found in prior work to primarily satisfy users' social and communication needs \cite{pelletier2020one,zhuSocialMediaHuman2015}. Since these platforms combine features like friends, likes, and reshares with algorithmic amplification, misinformation can spread rapidly, and people's trust in known contacts \cite{butler2023misinformation} can make them vulnerable. 

In addition to standard content moderation practices, such as removing, suspending, and downranking accounts or posts, governance strategies should leverage the close connections within these platforms by signaling account and message credibility \cite{kozyreva2024toolbox}. \textbf{Persistent credibility indicators} could label verified accounts with a proven history, since users on these platforms typically follow other accounts rather than content channels. \textbf{Sharing or searching friction}, such as accuracy nudges like "Have you read this?" when sharing unopened articles, and redirecting users to established sources or help lines when searching for harmful content, should also be introduced. Finally, \textbf{encouraging social corrections} is important because these platforms' higher proportion of close ties should allow them to facilitate social corrections through their design  \cite{kozyreva2024toolbox}.

For closed messaging platforms (WhatsApp), groups (Facebook), and broadcast stories (Instagram, Snapchat), approaches include \textbf{forwarding limits}, which cap the number of users a message can be forwarded to and have previously been used by WhatsApp \cite{porter2020whatsapp}, as well as \textbf{message-origin labels}, which let users see a message's origin and how many times it has been forwarded \cite{rossini2021dsyfunctional}. 


\subsection{Discourse Platforms}
Discourse platforms include profile-based and community-based platforms like LinkedIn \cite{tran2025me}, Nextdoor \cite{lillie2023nextdoor}, Reddit \cite{moore2017redditors}, as well as Facebook and X \cite{alhabash2024so,pelletier2020one}, the two platforms with the highest misinformation exposure in this study. These discussion-oriented platforms emphasize sharing, seeking opinions and information, and building communities, and their design encourages many-to-many engagement with topics and news, attracting users who actively seek out information and may be more willing to evaluate it. 

Governance strategies can leverage this community-oriented structure in several ways. They can establish effective \textbf{community reporting systems}, like X's "Community Notes" and Reddit's "Karma" points \cite{drolsbach2024community}. Another approach is \textbf{thread-level context panels} that display relevant context at the top of a thread, visible to users before they participate. Platform-level policies could \textbf{empower moderators} by setting platform minimum standards, such as requiring larger subreddits or threads to have increased community moderation (a form of governance that Reddit already has). Finally, platforms could enable users to query a \textbf{fact-checking agent} or bot integrated within the platform \cite{renault2026grok}.
These strategies empower communities to self-moderate, since users on these platforms may trust peers more than the platform.

\subsection{Discovery Platforms}
Discovery platforms serve as entertainment sources for entertainment, relaxation, and information, and include YouTube \cite{voorveldEngagementSocialMedia2018}, TikTok \cite{tran2025me}, Pinterest \cite{voorveldEngagementSocialMedia2018}, and Instagram, which has features that make it the only platform to encompass all three categories \cite{lu2022insta,pelletier2020one}. Usage tends to be passive and characterized by a one-to-many structure, with users discovering content related to their interests and hobbies rather than through a profile-based news feed, which in our study was associated with higher levels of self-reported misinformation exposure. Misinformation risk on these platforms stems mainly from recommendation algorithms rather than individual posts or users with large followings. 

Multiple governance strategies target recommendation algorithms and passive social media users \cite{kozyreva2024toolbox}. \textbf{Inoculation}, such as brief interstitials or media literacy nudges that play before a video starts, may be more effective for passive consumers than fact-checking labels. Platforms could be required to be more transparent about their algorithms or to conduct \textbf{periodic audits}. Finally, \textbf{creator-level credentialing} could be introduced with visible and verifiable creator badges for certain topics like health, science, or politics, instead of relying on standard account verification.
These approaches align with these platforms' content-focused nature, where content is recommended (unlike the direct interactions on interpersonal platforms or the topic-seeking on discourse platforms). 

\section{Conclusion}
This study analyzed how 1,010 US social media users perceive their exposure to and posting of misinformation across 11 platforms. Perceived exposure varies greatly by platform and is only moderately related to frequency of use: Facebook, not the most-used platform (YouTube), was where participants reported encountering the most misinformation, while closed platforms like WhatsApp and Snapchat showed higher exposure from known contacts. Mixed-effects models further indicated that both how often users engage and the platform itself predicted both reported exposure and posting behaviors. Gender and political ideology predicted exposure but not posting, and exposure itself was one of the strongest predictors of posting. These findings suggest a two-step process: some demographic factors may influence who is exposed or believes they are exposed to misinformation, while platform engagement determines who shares it.

Based on these results, we propose governance strategies tailored to three platform types (interpersonal, discourse, and discovery) since a governance tool's effectiveness likely depends on whether it aligns with the user's cognitive mode and purpose on that platform.
Platform-agnostic standards focused on the basic content and account moderation may serve as a minimum, but these governance frameworks should be continuously updated as platforms and usage evolve.


Several limitations nuance our work. Our sample focused on active US-based social media users who are more likely to be exposed to misinformation on social media platforms, but less active users might have different experiences. Since misinformation exposure and posting were self-reported, participants' perceptions of what counts as misinformation may not align with objective standards. 
Future research should explore exposure to and the posting of verified misinformation, and further examine how demographic factors influence perceptions of exposure to misinformation. 

\begin{credits}
\subsubsection{\ackname} 
The authors thank Samantha C. Phillips for her contribution to the survey’s design and data collection.
This work was supported by the Knight Foundation, the Office of Naval Research’s MURI: Persuasion, Identity, \& Morality in Social-Cyber Environments grant N00014-21-12749, Carnegie Mellon University’s Graduate small Project Help (GuSH), the Center for Computational Analysis of Social and Organizational Systems (CASOS), and the Center for Informed Democracy and Social-cybersecurity (IDeaS). The views and conclusions contained in this document are those of the authors alone. The funders have no role in study design, data collection and analysis, decision to publish, or preparation of the manuscript. 

\end{credits}

%
%
%
\bibliographystyle{splncs04}
\bibliography{references}

\end{document}